# One-step fabrication of metal nanostructures by high-throughput imprinting

Ze Liu

*Department of Engineering Mechanics, School of Civil Engineering, Wuhan University, Wuhan, Hubei, 430072, China. Email: ze.liu@whu.edu.cn*

**Direct nanoimprinting provides a simple and high-throughput route for producing uniform nanopatterns at great precision and at low costs. However, applying this technique to crystalline metals has been considered as impossible due to intrinsic limitation from grain size effect. Here we demonstrate direct superplastic nanoimprinting (SPNI) of crystalline metals well below their melting temperatures ($T_m$), generating ordered nanowire arrays with aspect ratio up to ~2000. Our investigations of replicating metal hierarchical nanostructures show the capability of imprinting features as small as 8 nm, far smaller than the grain size of bulk metals. Most surprisingly, the prepared metal hierarchical nanostructures were found possessing perfect monocrystalline structures. These findings indicate that nanoimprinting of crystalline metals below $T_m$ might be from lattice diffusion. SPNI as a one-step and highly controlled high-throughput fabrication method, could facilitate the applications of metal nanostructures in bio-sensing, diagnostic imaging, catalysis, food industry and environmental conservation.**

Controlled fabrication of metallic nanostructures plays a central role in much of modern science and technology [1-3] because the change in the dimensions of a nanocrystal enables tailor of its mechanic [4-6], electronic [7-10], optical [11-17], catalytic [2,18,19] and antibacterial properties [20]. It was found the properties of metal nanostructures significantly depend on their shapes and aspect ratios [15,20-22]. This has motivated an upsurge in research on the processing methods that allow better control of shape and size [20,23-26]. Chemical synthesis of metal nanoparticles has been well developed for preparing metal nanocrystals with good quality [27,28] since its first documented by Michael Faraday [29] but suffers from the limited selection of precursor compounds and has general challenge in dispersion of synthesized nanocrystals in liquids. The current preparation of metal nanopatterns mainly relies on advanced nanolithography techniques [3], such as nanosphere lithography [30,31], electron beam lithography [32,33]. These methods allow to fabricating homogeneously metallic nanopatterns but are costly owing to time-consuming, multistep processes and also limited in preparing nanostructures with low-aspect-ratios [34]. At present, high-throughput fabrication of metallic nanostructures in terms of controllability (e.g. resolution, precision, uniformity), material diversity, cost, and especially high-aspect-ratio remains a significant challenge [35-37].

Among the variety of developed nanofabrication methods, nanoimprinting [38-41], pioneered by Chou et al. [38], promises high-throughput fabrication of ordered and regular nanopatterns at great precision and at low costs but only limited in polymers [3,40,41] and a few of bulk metallic glasses (BMGs) with low supercooled liquid viscosities [39]. It has been generally considered infeasible to direct nanoimprinting of crystalline metals [35,39,42] because of the limitations on formability originating from fluctuations of plasticity on the nanoscale [43,44], size effects in plasticity [4,45], and grain size effect [42]. Here, we show direct nanoimprinting variety of crystalline metals (e.g. Bi, Ag, Au, Cu, Pt) by superplastic forming well below their melting temperatures. This technique enables one-step and rapid fabrication of metallic nanostructures with high precision and good controllability, in particular it allows to fabricating high-aspect-ratio nanostructures.

**Superplastic nanoimprinting of metals**

Figure 1a sketches the basic superplastic nanoimprinting steps for fabrication of crystalline metal nanostructures. A piece of metal is brought into contact with a mold, where the commonly used molds are $Al_2O_3$ templates prepared by anodic oxidation. SPNI is usually performed under temperature ranges of $0.5T_m \leq T < T_m$, with unit of absolute scale of temperature. Typically, after SPNI, the mold is dissolved to release the replicated metal nanostructures. Fig. 1b shows a typical as-thermoplastic formed sample under an optical microscopy (OM, 9XB-PC, Shanghai optical instrument factory), where a piece of Au was superplastic formed into a nanoporous $Al_2O_3$ template at 500 °C, well below the melting temperature of bulk Au ($T_m$ ~ 1064 ℃). The $Al_2O_3$ template was purchased from Hefei Pu-Yuan Nano Technilogy Ltd. with an average pore diameter of 55 nm. The uniform rust-red color of the sample in Fig. 1b suggests that Au nanowire arrays have been replicated, which is verified by characterizing the sample under a scanning electron microscopy (SEM, Zeiss Ultra Plus, Fig. 1c), after dissolving the $Al_2O_3$ template in KOH solution (concentration of 3 mol/L, temperature of 80 ℃).

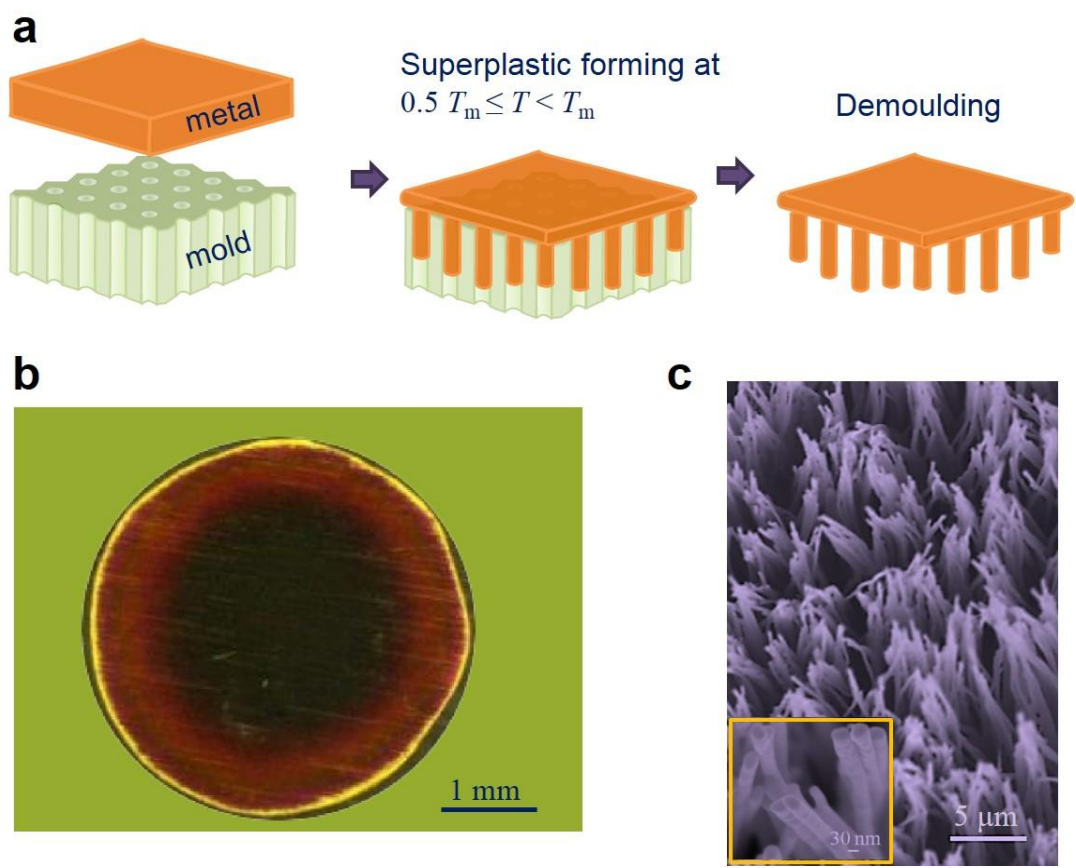

**Figure 1| Superplastic nanoimprinting of crystalline metals below melting temperatures. a**, Illustrative of the superplastic nanoimprinting process. **b**, Optical micrograph of an as-thermoplastic formed Au/$Al_2O_3$ template combination, which was prepared by SPNI under an applied force of 5 KN and holding for ~ 60 min. The uniform rust-red color of the sample suggests that Au nanowire arrays have been replicated, which is verified by characterizing the sample under SEM after dissolving the $Al_2O_3$ template in KOH solution (**c**).

**Metal nanowires with controlled aspect-ratio**

SPNI relies on direct mechanical deformation of imprint materials and can therefore achieve great replication precision in lateral dimensions. We now investigate the controllability of superplastic nanoimprinting along vertical direction. In general, the length of prepared metal nanowires by SPNI is a function of the forming pressure, processing temperature and time, which provides the way to control aspect ratio of replicated metal nanostructures. Taking nanoimprinting of Au as examples, we first investigated the effect of holding time on the aspect ratio of prepared Au nanowires, where $Al_2O_3$ templates with uniform cylindrical nanopore size of 200 nm were used. We first cut five Au short rods (61.8 ±0.5 mg) from an Au wire with diameter of 2 mm. Subsequently, we superplastic formed the prepared five Au rods into the $Al_2O_3$ templates at 500 ℃, under an applied force of 10 KN and held for 1, 4, 8, 20, 60 min, respectively. The length of Au nanowires at the center of each sample were measured under SEM, which has been used to calculate the aspect ratio of prepared Au nanowires (black dots in Figure 2a). Theoretically, we can generally quantify the length of replicated nanowires ($L$) under conditions of applied stress ($\sigma$) and temperature ($T$) as

$$L = f(\sigma, T, t, d) \tag{1}$$

where $d$ and $t$ are the diameter of nanopore and forming time, respectively. By applying classic Norton-Bailey's creep power law [46], we have

$$L = L_0 + A\exp\left(-\frac{Q}{RT}\right)\sigma^n t^m \tag{2}$$

where the constant $L_0$ approximates the length of metals flowing into the nanopores before load reaching the maximum force and it is in general a function of temperature, nanopore size and loading rate. $R$ and $Q$ are gas constant and active energy for creep, respectively. $A$ is a constant related to nanopore size and material properties such as elastic modulus, $n$ is the stress exponent of creep rate. In the above superplastic nanoimprinting experiments, the temperature is kept constant, the calculated mean forming pressure ($\bar{\sigma} = F_m/S$) for the five samples is $\bar{\sigma} = 471 \pm 24$ MPa (Extended Data Tab. 1), which can also be approximated as a constant. Thus the length and herein the aspect ratio of replicated nanowires simply obeys

$$L/d = B_1 + B_2 t^m \tag{3}$$

where $B_i$ ($i = 1, 2$) are constants related to nanopore size, material properties, processing temperature and forming pressure. Eq. (3) indicates that the aspect ratio of prepared metal nanowires continues to increase as the forming time increases, which can be adopted to fabricate high-aspect-ratio metallic nanostructures. This is very different with nanoimprinting of BMGs, where the processing time above glass transition temperature ($T_g$) is limited by the crystallization time due to the metastable nature of BMGs [39].

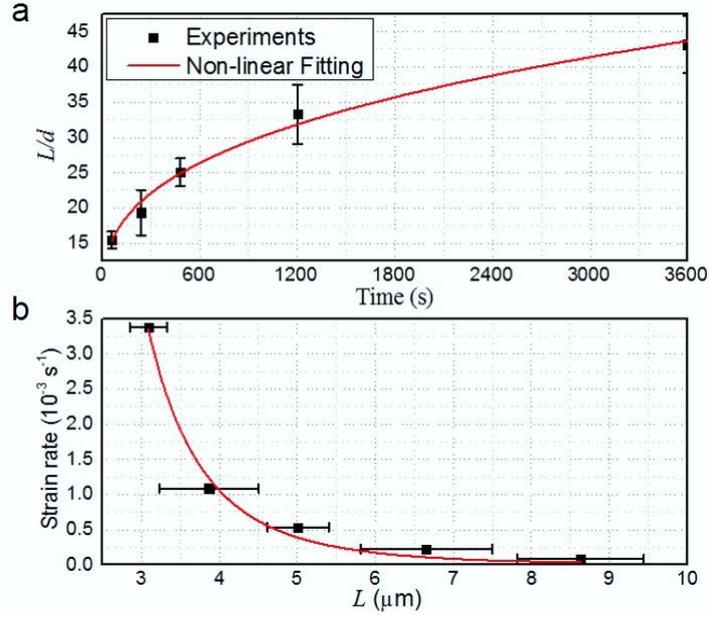

**Figure 2| Fabrication of Au nanowire arrays with controlled aspect ratios by varying the forming time during SPNI. a,** The length of Au nanowires *versus* holding time, where five Au short rods with diameter of 2 mm (61.8 ± 0.5 mg) were superplastic formed into 200 nm $Al_2O_3$ templates at 500 °C, under an applied force of 10 KN and holding for 1, 4, 8, 20, 60 min (black dots), respectively. The best fitting result by using eq. (3) gives $L/d = 6.34 + 2.17 \times t^{0.35}$ (red line). **b,** The calculated apparent strain rate ($\dot{\varepsilon}_{app} = d(\ln L)/dt$) in our experiments is on the order of $10^{-3}$ s$^{-1}$, and it is found scaling with $L$ as $\dot{\varepsilon}_{app} \propto L^{-4.45}$ by fitting with a power law (red line).

One of the doubts about direct nanoimprinting of crystalline metals is the expected high strain rates to activate superplasticity in metals [35]. However, according to eq. (3), the non-linear fitting our experimental data results in $L/d = 6.34 + 2.17 \times t^{0.35}$ (red line in Fig. 2a), which gives a very low apparent strain rate ($\dot{\varepsilon}_{app} = d(\ln L)/dt$), on the order of $10^{-3}$ s$^{-1}$ in our experiments (Fig. 2b), which is similar to the nanoimprinting of BMGs above $T_g$ [39]. Fig .2 b shows the calculated apparent strain rate versus the length of nanowires. It is obvious that $\dot{\varepsilon}_{app}$ decreases as the length of nanowire increasing (black dots in Fig. 2b), which is as anticipated since the longer of nanowires, the higher of flow resistance force from wall friction. Further fitting to the $\dot{\varepsilon}_{app}$ - $L$ data yields a power of -4.45 (red line in Fig. 2b), drastically deviating from the scaling law of Hagen–Poiseuille pipe flow, where $\dot{\varepsilon}_{app} \propto L^{-2}$ (see Methods section). Such a large discrepancy clearly rules out the viscous flow dominated mechanism in our SPNI experiments.

In addition to varying holding time to control the length of replicated Au nanowires, we also independently varied the processing temperature to study its effect on the aspect ratio of prepared Au nanowires (Extended Data Fig. 1). In this series of experiments, we firstly prepressed four Au short rods (29.5 ±0.5 mg) at 415 ℃ to get flat discs with thickness of 0.35 ± 0.05 mm. Subsequently we superplastic formed the Au flat discs into 100 nm $Al_2O_3$ templates at 415, 450, 496, 722 ℃, respectively, where all the samples were loaded to 3 KN and held for 100 s. The measured length of Au nanowires at the center of each sample is shown in Extended Data Fig. 1. It is clear that the length of nanowires increases with the rising of temperature, which can be understood from the Arrhenius-type temperature relation in eq. (2), the rising of temperature will increase the activity of atoms and/or defects.

**Replication of metal nanoarchitectures and TEM characterization**

At present, fabrication of metal nanostructures smaller than 10 nm is very challenging, even with advanced nanolithography techniques. To demonstrate the powerful of our SPNI technique to replicate extremely small features, we adopted a smallest nanomold we could get — a hierarchical $Al_2O_3$ template with multiply branched nanopores in its surface layer, where the branched nanopore size gradually increases from ~ 8 nm in the outmost surface to 200 nm in the base layer (Extended Data Fig. 2). Figure 3a shows SEM images of replicated Au hierarchical nanostructures by using the hierarchical $Al_2O_3$ template (see Methods section). The aggregation of Au nanostructures into bundles makes it difficult to examine the replicated hierarchical structures. We therefore transferred the prepared Au nanostructures onto a transmission electron microscope (TEM) mesh grid (see Methods section). SEM imaging of Au hierarchical nanostructures on the TEM mesh grid clearly shows the primary stems abruptly multiplied by several small branches (Fig. 3b). The replicated smallest branch is ~ 8 nm (Extended Data Fig. 3), in consistent with the hierarchical $Al_2O_3$ template we used, indicating the high replicating fidelity of our SPNI technique.

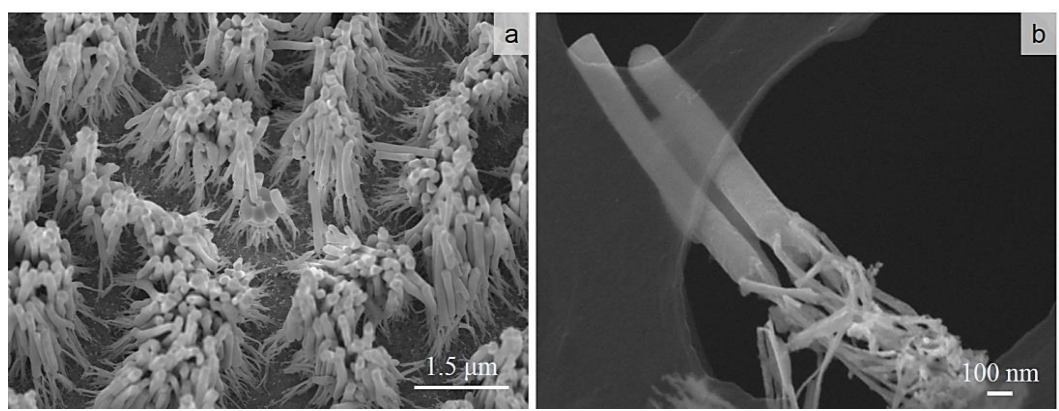

**Figure 3| Au hierarchical nanostructures.** Au hierarchical nanostructures were replicated by superplastic forming a piece of Au into a hierarchical $Al_2O_3$ template with multiply branched nanopores in its surface layer at ~ 500 ℃, under an applied force of 5 KN and holding for ~ 10 min. **a**, SEM micrograph of fabricated Au hierarchical nanostructures. **b**, Zoom-in imaging of Au

hierarchical nanostructures on a TEM mesh grid (see Method section) clearly shows the primary stems abruptly multiplied by several small branches.

It is particularly noteworthy that the nanopore size at the entrance of the hierarchical $Al_2O_3$ template is only ~ 8 nm, three orders of magnitude smaller than the grain size of the bulk Au we used ($10^1$ μm, Extended Data Fig. 4), which suggests that the successful replication of metal nanostructures should originate from creep deformation of single crystals because most nanopores in the $Al_2O_3$ template are in contact with single crystals during SPNI. Considering that nanopores in the template will block the propagation of dislocations, called nanoscale geometrical confinement [4], the usual dislocation slip and twining mechanisms cannot work here. Direct evidence is from the characterization of Au hierarchical nanostructures under a high-resolution TEM (JEM2100F, see Methods section) — there is no dislocations in the interior of replicated Au nanostructures and no observed slip steps on their surfaces (Figure 4). Fig. 4c shows a typical diffraction pattern for the selected Au nanostructure in Fig. 4b, which exhibits symmetrical and clear electron diffraction spots for single face-centered cubic (fcc) crystal. The growing axis of the single crystal is determined along <111> crystallographic orientation. High-resolution TEM images (Fig. 4d-g) and fast Fourier transformations (insets of Fig. 4d-g) at the selected regions denoted by A, B, C, D in the nanostructure in Fig. 4b confirm its perfect single-crystallinity again. Most surprisingly, we observed that even around the regions where the branches abruptly converged (Fig. 4e and f), the crystallinity is not disturbed at all.

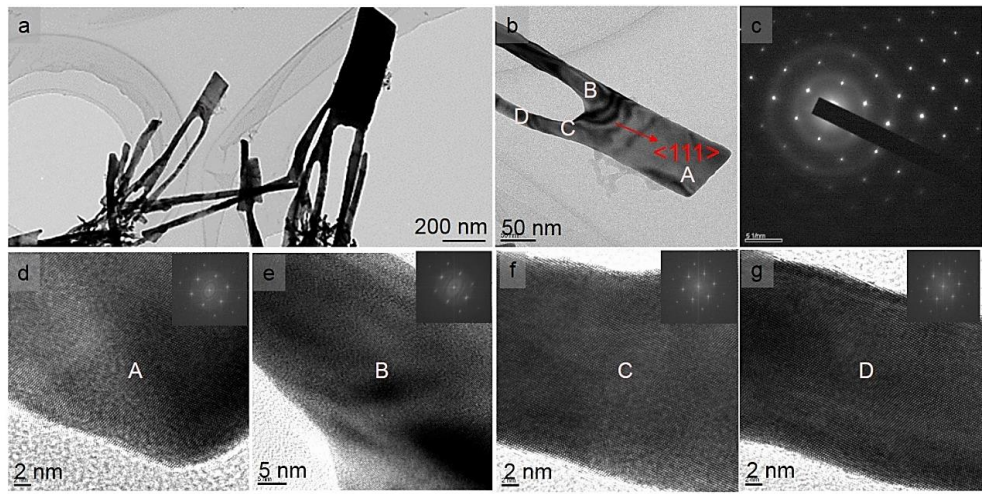

**Figure 4| Characterization of Au hierarchical nanostructures by using transmission electron microscope. a-b**, Topography images of prepared Au nanostructures. **c**, Diffraction pattern of the Au hierarchical nanostructure in (**b**) showing a fcc single crystal structure and the axis of the nanostructure is determined along <111> crystallographic orientation. **d-g**, High-resolution TEM images at the regions denoted by A, B, C, D in (**b**) and fast Fourier transformations (insets) confirming the perfect single crystal of the Au hierarchical nanostructure again.

Among the typical creep deformation mechanisms of viscous flow, dislocation motion, lattice or grain boundary diffusion, dislocation motion is attributed to the main creep mechanism of crystalline metals at high temperatures, i.e. T > $0.4T_m$ [47]. In view of the fact that most of metals possess grain sizes larger than 1 μm, direct nanoimprinting of crystalline metals below $T_m$ has been considered as impossible [35,39,42]. However, we noted that both viscous flow and lattice diffusion are independent of grains, which provides the possible mechanisms for direct nanoimprinting of metals. Considering that our investigations of the effect of holding time on length of replicated nanowires has ruled out the viscous flow dominated mechanism as discussed before (Fig. 2), it is quite possible that the deformation mechanism of superplastic nanoimprinting crystalline metals below $T_m$ originates from lattice diffusion.

**Fabricating variety of metal nanostructures by SPNI**

Our SPNI technique includes but not limited to fabricating Au nanostructures. To show the generalization of this SPNI technique for preparing crystalline metal nanostructures, we also fabricated Bi, Ag, Cu, and Pt nanowire arrays (Figure 5). Fig. 5a shows the fabricated high-aspect-ratio Bi nanowire arrays with aspect ratio of AR ~ 300, which is achieved by superplastic nanoimprinting a piece of Bi at 260 ℃, closing to its melting temperature ($T_m$ ~ 273 ℃). We observed that Bi can completely fill the template within 36 s under the applied force of 8 KN. Beside increasing processing temperatures, high-aspect-ratio metal nanowires can also be fabricated through increasing the forming time as suggested by eq. (3). Based on which we replicated Ag nanowire arrays with an extremely high aspect ratio of ~ 2000 (Fig. 5b) since the Ag nanowires almost fill up the $Al_2O_3$ template during the SPNI (thickness of the $Al_2O_3$ template ~ 50 μm, Extended Data Fig. 5). Additionally, we also show in Fig. 5c-d for replicated Cu and Pt nanorod arrays. It is noted that bulk Pt possesses melting temperature as high as 1772 ℃, which can also be nanoimprinted at ~ 820 ℃ by using our method (Fig. 5d). The clear crystal facets and regular shapes observed at the top of Pt nanorods shows the superior of SPNI again for replicating metal nanostructures with excellent crystallinity.

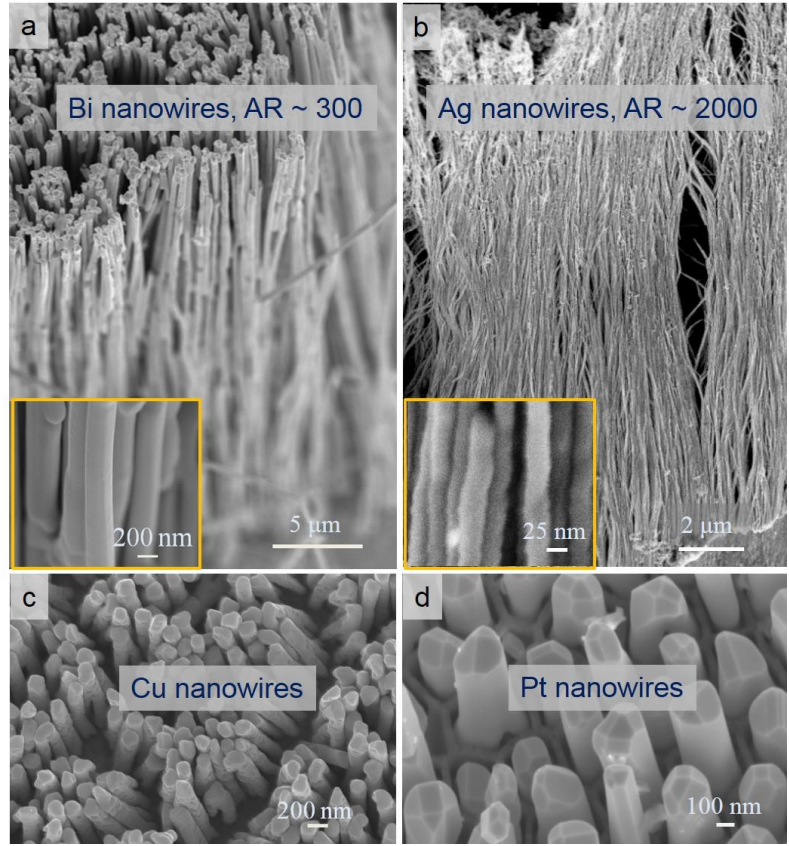

**Figure 5| Scanning electron microscopy images of Bi, Ag, Cu, Pt nanowire arrays. a**, Superplastic nanoimprinting a piece of Bi by using 200 nm $Al_2O_3$ template at 260 ℃ (closing to its melting temperature, $T_m$ ~ 273 ℃) and under a force of 8 KN. Bulk Bi completely filled the $Al_2O_3$ template within 36 s, corresponding to an aspect ratio of ~300 since the thickness of $Al_2O_3$ template is ~ 60 μm. **b**, An extremely high-aspect-ratio of ~ 2000 for Ag nanowires was also obtained by superplastic nanoimprinting a piece of Ag into 25 nm $Al_2O_3$ template at ~ 700 ℃, under an applied force of 15 KN and holding for 90 min. **c-d**, Cu and Pt nanowire arrays were fabricated by superplastic nanoimprinting at ~ 550 and ~ 820 ℃, respectively.

**Surface-enhanced Raman scattering from Au nanowires**

Reproducible and robust metal nanostructures that strongly enhance the electromagnetic field are most desirable for surface-enhanced Raman scattering (SERS) but are difficult to achieve [36,48]. We have shown above that direct thermoplastic nanoimprinting of crystalline metals offers a rapid and controllable method to fabricate uniform metallic nanostructures, which provides an ideal way to fabricate robust SERS substrates. We now explore how effective thermoplastic nanoimprinted metal nanowire arrays as SERS substrates. As examples, we prepared five Au flat discs, four of them were imprinted with Au nanowire arrays by SPNI and the remaining one is used as a reference sample (see Methods section). For simplicity, we denote the reference sample as bulk Au and the four samples attached with Au nanowire arrays as s1, s2, s3 and s4, corresponding to nanowire size of 200, 90, 55, 20 nm, respectively. The optical micrographs of the as-thermoplastic formed Au/Al$_2$O$_3$ template combinations show clearly size-dependent colors (Extended Data Fig. 6a-d): the surface color changes from pinkish (s1), rust red (s2) to olive-green (s3) and finally tends to grey (s4). Such a absorption redshift is well understood from the size-dependent surface plasmon oscillation [21] and it indicates the effectiveness of the prepared samples as SERS substrates.

Extended Data Fig. 6i shows measured Raman signals at the center of the prepared five samples by using $1.0 \times 10^{-5}$ M crystal violet (CV) as a sensitive SERS analyte to detect the electromagnetic enhancement (RENISHAW Raman microscope, INVIA, see Methods section). Although CV molecular absorbed on the bulk Au shows rather weak Raman signals, five Raman lines located at 442 cm$^{-1}$, 802 cm$^{-1}$, 1174 cm$^{-1}$, 1384 cm$^{-1}$ and 1620 cm$^{-1}$ can still be recognized and these Raman shifts agree well with literatures report (Extended Data Tab. 2). On the contrary, almost all of the reported Raman lines for CV between 400 and 1800 cm$^{-1}$ are drastically intensified by samples s1-s4 (Extended Data Fig. 6i and Extended Data Tab. 2), which demonstrates that our SPNI technique is very suitable to fabricate metallic SERS substrates. Extended Data Fig. 6i also demonstrates that the Raman shifts decrease as the size of nanowires increased from 90, 200 nm to infinity (bulk Au). Such a size dependent SERS signals is in good agreement with theoretical prediction [21,22]. However, when the size of nanowires continues to decrease below 90 nm, the Raman signals become slightly weaker rather than continuous increasing, which we attribute to the aggregation of small nanowires to form big bundles due to mechanical instability (Extended Data Fig. 6g-h).

**Conclusions**

Direct superplastic nanoimprinting of crystalline metals well below $T_m$ has enabled us to fabricate variety of metal nanowires with aspect ratio up to ~ 2000. By adopting a hierarchical Al$_2$O$_3$ template with gradient nanopore size, we are able to replicate Au hierarchical nanostructures. The perfect monocrystalline structure of the prepared Au hierarchical nanostructures, together with the fact that the nanopore size at the entrance of the Al$_2$O$_3$ template is only ~8 nm, far smaller than the grain size of the bulk Au, we argue that direct superplastic nanoimprinting of crystalline metals might originate from lattice diffusion dominated mechanism. Finally, we show our SPNI technique is very suitable to fabricate metallic substrates for SERS application. SPNI is inherently high-throughput due to its parallel printing, and it requires only one step and simple equipment set-up, leading to low-cost. We propose our technique should facilitate the applications of metal nanostructures in catalysis, nanoelectronics, sensors and plasmonics.

**Acknowledgements**

This research was supported by the start funding from Wuhan University, the Fundamental Research Funds for the Central Universities, Hubei Provincial Natural Science Foundation of China through Grant 2016CFB159 and National Natural Science Foundation of China through Grant 11602175.

Correspondence and Requests for materials should be addressed to: Dr. Ze Liu (ze.liu@whu.edu.cn)